\documentclass[prl,aps,twocolumn,nofootinbib,,superscriptaddress]{revtex4-1}
\usepackage{amsmath}
\usepackage{amsfonts}
\usepackage{amssymb}

\usepackage{graphicx}
\usepackage{dcolumn}
\usepackage{bm}
\usepackage{color}

\begin{document}
\title{Strain-induced partially flat band, helical snake states, and interface superconductivity in topological crystalline insulators}

\author{Evelyn Tang}
\affiliation{Department of Physics, Massachusetts Institute of Technology, Cambridge, MA 02139}
\author{Liang Fu}
\affiliation{Department of Physics, Massachusetts Institute of Technology, Cambridge, MA 02139}

\begin{abstract}

\end{abstract}

\maketitle

{\bf Topological crystalline insulators in IV-VI compounds host novel topological surface states consisting of multi-valley massless Dirac fermions at  low energy. Here we show that strain generically acts as an effective gauge field on these Dirac fermions and creates pseudo-Landau orbitals without breaking time-reversal symmetry. 
We predict the realization of this phenomenon  in IV-VI semiconductor heterostructures,
due to a naturally occurring misfit dislocation array at the interface that produces a periodically varying strain field. 
Remarkably, the zero-energy Landau orbitals form a flat band in the vicinity of the Dirac point, and coexist with a network of snake states at higher energy. We propose that the high density of states of this flat band gives rise to interface superconductivity observed in IV-VI semiconductor multilayers at unusually high temperatures, with non-BCS behavior. Our work demonstrates a new route to altering macroscopic electronic properties to achieve a partially flat band, and paves the way for realizing novel correlated states of matter. }

The recently discovered topological crystalline insulators host novel topological surface states that are protected by the symmetry of the underlying crystal\cite{hsieh, ando, poland, hasan}.
At low carrier energy, these surface states consist of multi-valley massless Dirac fermions, whose characteristic properties are highly tunable by external perturbations. Breaking the crystal symmetry at the atomic scale generates a Dirac mass and leads to gapped phases\cite{hsieh, okada} with potentially novel functionalities in low-power electronics and spintronics\cite{liuhsieh, fang, zhang}. 

The (001) surface states of topological crystalline insulators SnTe and Pb$_{1-x}$Sn$_x$Te(Se) consist of massless Dirac fermions 
at four valleys that exhibit spin texture of the same chirality\cite{hsieh}. The four Dirac points are located at two pairs of opposite momenta, denoted by $\pm {\bf K}_1$ and $\pm {\bf K}_2$, in the vicinity of the $X_1$ and $X_2$ point in the surface Brillouin zone respectively. ${\bf K}_1$ and ${\bf K}_2$ are related by four-fold rotation around the surface normal, while $+{\bf K}_j$ and $- {\bf K}_j$ are related by time-reversal symmetry (see Fig.1a). 
Importantly, unlike in the case of topological insulators\cite{kane, qi, jmoore}, the Dirac points in topological crystalline insulators are not pinned at time-reversal-invariant momenta\cite{liu}, because their massless nature is protected by crystal symmetry instead of time reversal\cite{fu, moore}.  As a consequence, perturbations can move such Dirac points in momentum space, mimicking the effect of a  gauge field vector potential  without breaking time reversal symmetry. 
Two effective ways of moving surface Dirac points in topological crystalline insulators are (i) alloy composition tuning, as recently demonstrated in Pb$_{1-x}$Sn$_x$Te(Se)\cite{yando} and (ii)  strain, which is the subject of this work.  

\begin{figure*}[tb]
\includegraphics[width=0.9\linewidth]{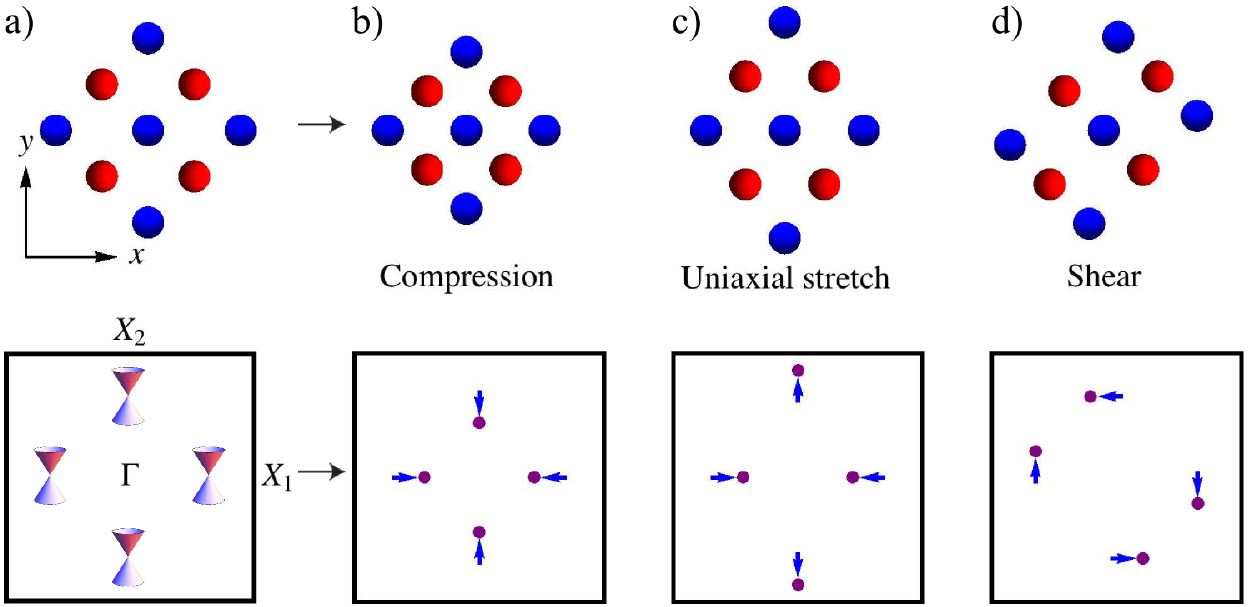}
\caption{{\bf Strain-induced Dirac point shift on the (001) surface of a topological crystalline insulator such as SnTe and Pb$_{1-x}$Sn$_x$Se(Te).} Top: strain on the rocksalt structure of a) compression, b) uniaxial stretch and c) shear; the $x$ and $y$ axes are in the (110) and (1$\bar1$0) directions respectively. 
Bottom: the corresponding strain-induced shift (arrows) of Dirac points to new positions (circles) in the (001) surface Brillouin zone, which is equivalent to an effective gauge field. }\label{fig:strain}
\end{figure*}

We first use symmetry analysis to determine  the general form of strain-induced Dirac point displacement on the topological crystalline insulator (001) surface, 
which is equivalent to a strain-induced gauge field vector potential ${\bf A}_j \equiv {\bf K}_j' - {\bf K}_j$ acting on the Dirac fermion at valley ${\bf K}_j$. 
To lowest order, ${\bf A}_j = (A_j^x, A_j^y)$ is linearly proportional to the strain field $u_{ij}$ as given by (see Methods): 
\begin{eqnarray}
{\bf A}_1 &=&  
(  \alpha_1 u_{xx} + \alpha_2 u_{yy}  ,  \;   \alpha_3 u_{xy}  ) ,  \nonumber \\
{\bf A}_2 &=&  
(\alpha_3 u_{xy} ,  \; \alpha_1 u_{yy} + \alpha_2 u_{xx} ).  \label{a}
\end{eqnarray}
where $u_{ij} \equiv (\partial_j u_i + \partial_i u_j)/2$ is the spatial gradient of the displacement field $\bf u$, and the coordinate axes $x$ and $y$ 
are along the [110] and [$\bar{1}$10] directions respectively. $\alpha_{1},\alpha_2$ and $\alpha_3$ denote  three independent coupling constants. 
As shown in Fig.1, under a given type of strain, the Dirac fermions at valleys ${\bf K}_1$ and ${\bf K}_2$ experience distinct gauge fields ${\bf A}_1$ and ${\bf A}_2$. In addition, opposite gauge fields $-{\bf A}_j$ are induced at the Dirac valleys $-{\bf K}_j$ as required by time reversal symmetry.

A uniform strain can only produce a constant gauge field vector potential, which yields zero pseudo-magnetic field ${\bf B} \equiv \nabla \times \bf A$. 
Instead, a highly inhomogeneous strain is required to create a strong pseudo-magnetic field, which is difficult to engineer with high precision and control. 
Such a field was previously observed in graphene nanobubbles\cite{levy, manes, guinea, neto}, which are however localized 
within nanoscale regions thus leaving electronic properties on a large scale essentially unaltered.    

\begin{figure}[tb]
\begin{center}
\includegraphics[width=\linewidth]{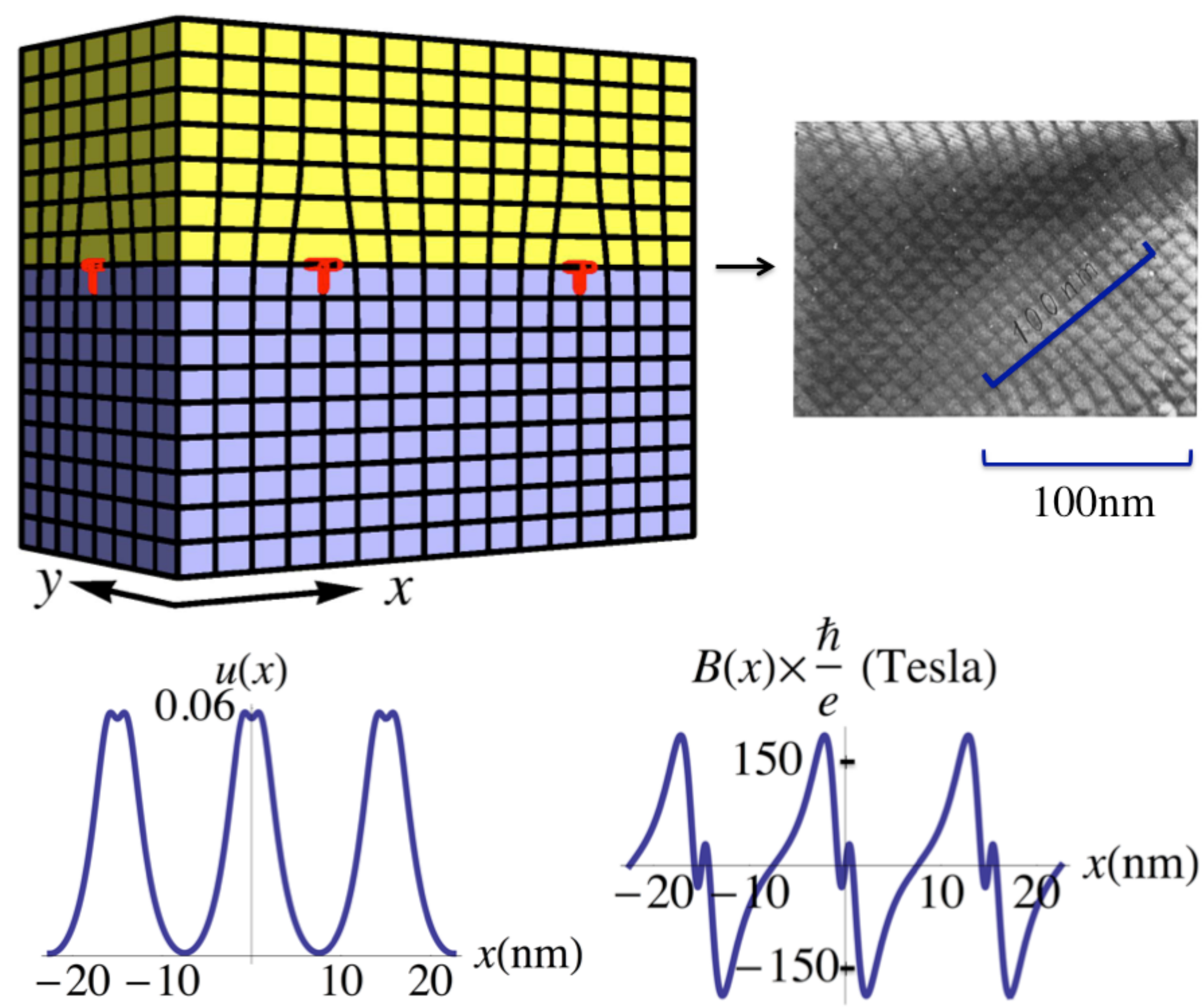}
\caption{{\bf The spontaneous formation of a misfit dislocation array gives rise to a periodically varying strain field and pseudo-magnetic field, at the interface.}
Top:   a square array of misfit edge dislocations is spontaneously formed 
at the (001) interface of two IV-VI semiconductors (e.g., PbTe/PbSe), due to lattice mismatch. 
The array consists of dislocation lines along both $x$ and $y$ directions, as shown in the transmission electron microscopy image taken from Ref.\cite{dis-sc}.  
Bottom: the set of dislocation lines along the $y$ direction creates a periodically varying strain field $u(x) \equiv u_{xx}(x) + u_{yy}(x)$ as a function of $x$ given by Eq.(\ref{uyy}). This is plotted here using the realistic parameters: $\lambda=15$nm, $z=2$nm, $\nu=0.26$ and $a=6.4\AA$ (see main text), together with the pseudo-magnetic field $B(x)$  it generates.
}\label{fig:misfit}
\end{center}
\end{figure}

Here we show that a periodic pseudo-magnetic field covering macroscopic spatial regions arises naturally in IV-VI semiconductor (001) heterostructures consisting of a topological crystalline insulator and a trivial insulator. This field is created by misfit dislocations that are known to spontaneously form at the interface due to lattice mismatch. 
Remarkably,  these dislocations self-organize into a nearly perfect two-dimensional square array with a period of 3-25nm (see Fig.2), as observed by transmission electron spectroscopy, X-ray diffraction, and scanning tunneling microscopy in SnTe/PbTe, PbTe/PbS, PbTe/PbSe, and PbTe/YbS\cite{med, med2, springholz}. 
This  dislocation superstructure naturally produces an inhomogeneous  strain field, which gives rise to an unusual pseudo-magnetic field ${\bf B}(x,y)$ that
alternates with a nanoscale period and averages to zero. Instead, it should be pointed out that the widely studied  uniform pseudo-magnetic field is physically impossible in the thermodynamic limit, due to the bounded nature of strain.

To calculate ${\bf B}(x,y)$, we first note the experimental observation\cite{springholz} that the dislocation array is a superposition of  two sets of equally spaced 
parallel dislocation lines along the $x$ and $y$ directions respectively. Each line is an edge dislocation with a Burgers vector ${\bf b} = \frac{a}{2} \hat{z} \times {\hat e}$ ($a$ is the lattice constant), which is 
parallel to the interface and perpendicular to the line direction ${\hat e}= \hat x$ or $\hat y$.
The set of dislocation lines along the $y$ direction creates a displacement field at the interface. The corresponding strain field contains two in-plane components  $u_{xx}$ and $u_{yy}$ that periodically vary in the $x$ direction. $u_{xx}$ and $u_{yy}$ are the sum of strains from each dislocation line indexed by $N$:  
\begin{eqnarray}
u_{xx}(x) &=& \sum_N u^0_{xx}(x-N\lambda), \nonumber \\
u_{yy}(x) &=& \sum_N u^0_{yy}(x-N\lambda), 
\end{eqnarray}
where $\lambda$ is the dislocation array period. 
Within linear elasticity theory, $u^0_{xx}$ and $u^0_{yy}$  are given by\cite{lubensky}  
\begin{eqnarray}
u^0_{xx} (x) &=& \frac{ b z}{2\pi (1-\nu)}  \frac{(3 x^2 + z^2)}{(x^2+z^2)^2},  \nonumber \\
u^0_{yy} (x) &=&  \frac{ b z \nu}{\pi (1-\nu)} \frac{1}{x^2 + z^2}. 
\label{uyy}
\end{eqnarray}
Here $z$ is the distance from the dislocation plane, $\nu$ is the Poisson's ratio and $b$ is the magnitude of the Burger's vector ${\bf b}$.
Similarly, the set of dislocation lines along the $x$ direction creates strain fields $\tilde{u}_{xx}$ and $\tilde{u}_{yy}$, 
which are related to $u_{xx}$ and $u_{yy}$ by the $\pi/2$ rotation: 
$\tilde{u}_{xx}(y) = u_{yy}(x\to y)$ and $\tilde{u}_{yy}(y) = u_{xx}(x\to y)$.

The total strain fields $u_{xx}(x) + \tilde{u}_{xx}(y)$ and $u_{yy}(x) + \tilde{u}_{yy}(y)$ create gauge fields for the two-dimensional Dirac fermions at the interface. 
It follows from Eq.(\ref{a}) that  the gauge field ${\bf A}_j$ for the Dirac fermion at ${\bf K}_j$ contains both longitudinal and transverse components, 
${\bf A}_j = {\bf A}_j^L + {\bf A}_j^T$. The longitudinal component ${\bf A}_j^L$ can be ``gauged away'' by a unitary transformation hence will not be considered below.  
The transverse component ${\bf A}_j^T$ is given by 
\begin{eqnarray}
{\bf A}^T_1( y) &=&  (\alpha_1 \tilde{u}_{xx}(y) + \alpha_2 \tilde{u}_{yy} (y)  , 0 ) , \nonumber \\
{\bf A}^T_2 (x) &=& ( 0 ,  \alpha_1 u_{yy}(x) + \alpha_2  u_{xx}(x) ),  \label{at}
\end{eqnarray}
which yields an out-of-plane pseudo-magnetic field $B_{j}$ acting on the Dirac fermion at valley ${\bf K}_j$: $B_1 (y) = \nabla \times {\bf A}^T_1(y)$ 
and $B_2 (x) = \nabla \times {\bf A}^T_2(x)$.   
Note that although the dislocation array is two-dimensional, the pseudo-magnetic field for a given Dirac valley is periodically alternating in one direction only.  

We now estimate the magnitude of the pseudo-magnetic field created by misfit dislocation arrays at the interface.   
For a typical array period of $\lambda=15$nm at $z=2$nm, with the Poisson ratio for PbTe of $\nu=0.26$\cite{weber} 
and the lattice constant $a = 6.4\AA$, the corresponding strain field $u = u_{xx} + u_{yy}$, plotted in Fig.\ref{fig:misfit},  has  
a maximum value of $6\%$, which is comparable to the $3-10$\% lattice mismatch in IV-VI semiconductor heterostructures\cite{med, med2, springholz}. 
A recent ab-initio calculation\cite{strain1} finds that the Dirac point positions of strained PbTe in the topological crystalline insulator phase shift linearly 
under compression, and yields $\alpha_1=2.2\AA^{-1}$. Assuming $\alpha_2 \approx \alpha_1 \equiv \alpha$,  we find a  Dirac point displacement of $A_0 = 0.13\AA^{-1}$ 
under the maximum 6\% strain, which is comparable to the displacement under changes in alloy composition in Pb$_{1-x}$Sn$_x$Te as recently observed  by angle-resolved photoemission spectroscopy\cite{yando}.  
Using this value of $\alpha$, we plot the periodically alternating pseudo-magnetic field created by the dislocation array  in Fig.\ref{fig:misfit}.  
The maximum field strength is around 180T.

How does this pseudo-magnetic field  change the electronic structure of Dirac fermion surface states at the interface? 
We first analyze this problem using a local field approximation. 
When the magnetic field is uniform,  two-dimensional massless Dirac fermions form a set of Landau levels.   
In the Landau gauge, the Landau orbitals are one-dimensional strips localized at different  $x$ positions and infinite in the $y$ direction. 
The width of the strip is set by the magnetic length $L_B=1/\sqrt{|B|}$. 
When the magnetic field $B(x)$ is slowly varying over the distance $L_B$,  the Landau level strip remains an approximate energy eigenstate. 
However, the Landau level energy becomes position dependent and is determined by the local magnetic field: 
\begin{eqnarray}
E_n(x) = {\rm sgn} (n) \sqrt{ 2 n v_x v_y |B(x)|}.   \label{ex}
\end{eqnarray}    
where $v_{x}(v_{y})$ is the Dirac fermion velocity in the $x (y)$ direction.  
Since the $x$ position of a Landau level strip is proportional to its momentum in the $y$ direction i.e. $x\propto {\rm sgn} (B)k_y$, the Landau level energy $E_n(x)$ as a function of 
position $x$  is also the dispersion as  a function of momentum $k_y$. 
Thus, when the magnetic field is spatially varying, Landau levels at different positions become non-degenerate and collectively form a {\it dispersive} band. 
It should be clear from this analysis that  the extensive degeneracy at $E=0$  holds so long as the strain field varies slowly over $L_B$ even without being strictly periodic. 

  
The $n=0$ Landau level of massless Dirac fermions deserves special attention. Its energy is pinned at $E_0=0$ independent of the magnetic field strength, as recently observed in topological crystalline insulators\cite{okada}.  This implies that in a slowly varying pseudo-magnetic field,  the $n=0$ Landau orbitals     
remain extensively degenerate at zero energy, forming a flat band.   

To demonstrate this, we numerically calculate the energy spectrum of massless Dirac fermions 
under the periodically alternating pseudo-magnetic field created by the dislocation array. 
The Hamiltonian is given by   
\begin{eqnarray}
 H =  - i v_x \partial_x s_y - v_y (k_y -A_y(x) )s_x, \label{h}
\end{eqnarray}
where $s_x$ and $s_y$ are Pauli matrices. Here we have chosen to study the Dirac valley ${\bf K}_2$, for which the strain-induced gauge field $A_y(x)$ is given in Eq.(\ref{at}). 
For simplicity,  we approximate the strain field shown in Fig.\ref{fig:misfit} as a cosine function in our calculation, i.e., $A_y(x) = A_0 \cos (2\pi x/ \lambda)$.  
The main findings presented below are independent of the specific choice of $A_y(x)$.

It is instructive to express $H$ in terms of the dimensionless quantities $\tilde{x} \equiv x /\lambda$, $\tilde{y} \equiv  v_x y / v_y \lambda$ and $\tilde{H} \equiv  \lambda H / v_x$: 
\begin{eqnarray}
\tilde{H} = - i \partial_{\tilde{x}} s_y - (\tilde{k}_y - \beta \cos(2\pi \tilde{x})) s_x.  
\end{eqnarray}
The energy spectrum of $\tilde{H}$ is entirely determined by a single dimensionless parameter $\beta =  (v_y/v_x) A_0 \lambda \propto \sqrt{\lambda/L_B}$, 
which depends on the ratio of the dislocation array period and the magnetic length. 
The formation of local Landau levels and flat bands  requires  $\beta \gg 1$. 
To estimate the realistic value of $\beta$, we use the aforementioned parameter $A_0=0.13\AA^{-1}$, and take
 Dirac velocities from an ab-initio calculation\cite{liu}: $v_x= 0.84 eV\AA, v_y=1.3eV\AA$.  
This yields $\beta \approx 20-40$ for $\lambda=10-20$nm, fulfilling the condition for flat bands.  
 
\begin{figure}
\begin{center}
\includegraphics[width=\columnwidth]{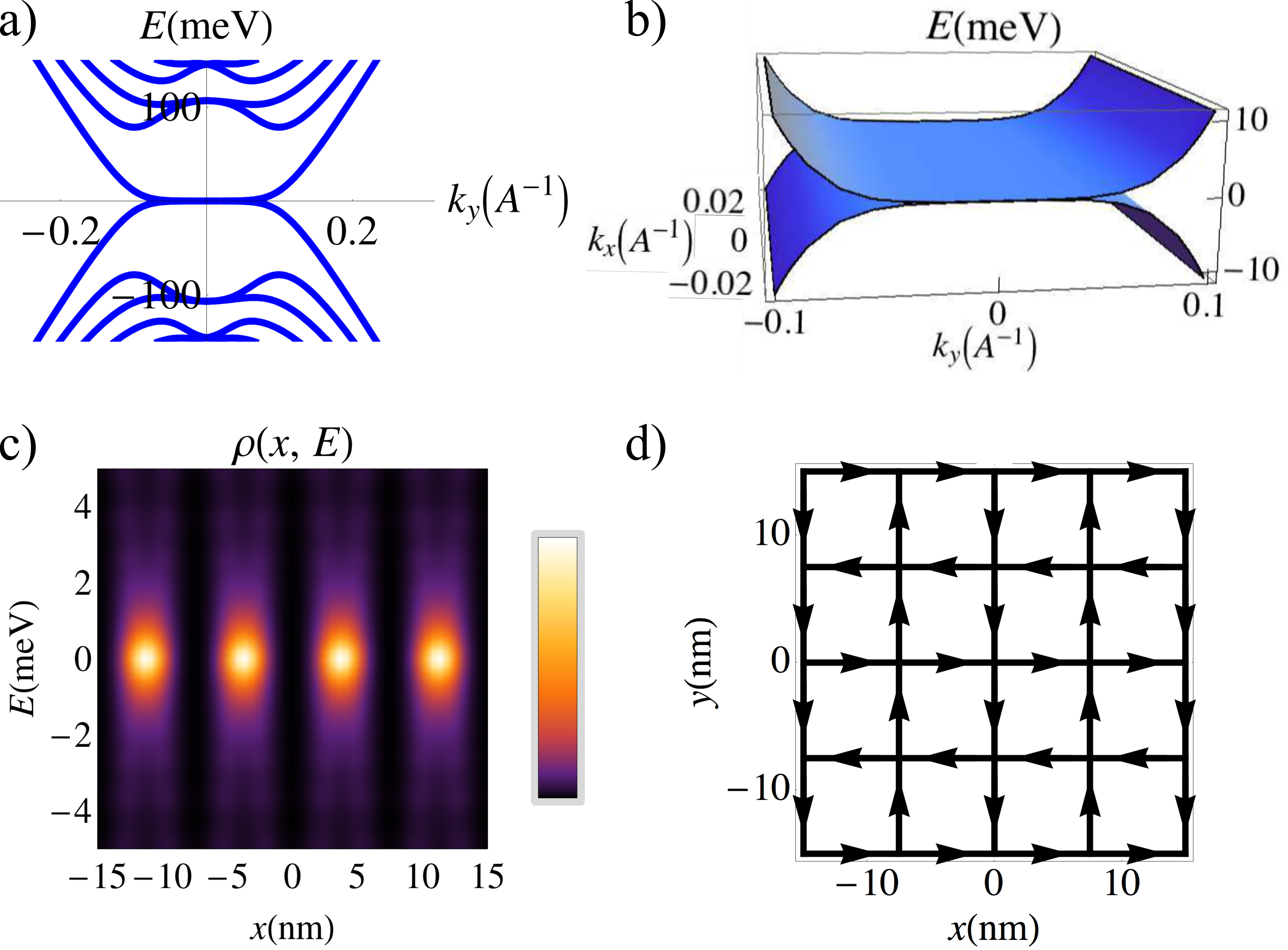}
\caption{{\bf The pseudo-magnetic field from strain creates flat bands and snake states.} (a) Under a periodically alternating pseudo-magnetic field $B(x)$, the initially linear Dirac dispersion becomes flat within a finite range of $k_y$ in the vicinity of the Dirac point, while higher Landau levels are more dispersive. (b) The band is completely flat in $k_x$, which has a much reduced period given by the dislocation superlattice Brillouin zone.
 (c) Local density of states $\rho(x,E)$ as a function of position $x$ and energy $E$, showing zero-energy Landau levels from regions of both positive and negative pseudo-magnetic field, which are spatially separated by dispersive snake states. (d) One-dimensional snake states appear where the pseudo-magnetic field changes sign. This schematic cartoon  shows snake states from valleys $ {\bf K}_1$ and $ {\bf K}_2$ moving along the $y$ and $x$ directions respectively, which form a two-dimensional network. Please refer to the main text for parameters used. }
\label{fig:bandswf}
\end{center}
\end{figure}

A representative band structure of $H$ for $\beta=30$ is plotted in Fig. \ref{fig:bandswf}a. As expected, we find two bands become very flat at zero energy for $|k_y|<k_c =0.05\AA^{-1}$. 
These two degenerate flat bands correspond to two types of $n=0$ Landau level strips that reside  in different 
spatial regions under positive and negative pseudo-magnetic fields respectively.  
The $x$ position of  Landau level strips in $B>0$ ($B<0$) regions increases (decreases) linearly with $k_y$.   
For $|k_y|>k_c$, these two types of  Landau levels become hybridized in ``transition regions'' where $B$ goes through zero, which are centered 
either on a dislocation line at $x = N\lambda$ or at the midpoint $x = (N +\frac{1}{2})\lambda$ between adjacent dislocations. 
As a consequence of  the level splitting from hybridization,  
the two zero-energy flat bands at $|k_y|<k_c$ evolve into a pair of dispersive bands at $|k_y| >k_c$, which 
reside on the domain wall between $n=0$ Landau levels under positive and negative pseudo-magnetic fields (see Fig. \ref{fig:bandswf}b).     

These dispersive states have a topological origin related to the unique 
half-integer Hall conductance of the $n=0$ Landau level of a single massless Dirac fermion: 
$
\sigma_{xy}  = {\rm sgn} (\mu B) \frac{1}{2} \frac{e^2}{h}, \label{xy}
$
where $\mu$ is the chemical potential. Because Hall conductances in $B>0$ and $B<0$ regions differ by $\Delta \sigma_{xy} = {\rm sgn}(\mu) e^2/h$,  
there exists one branch of chiral snake states at the domain wall\cite{network}, which is derived from the valley ${\bf K}_2$ and moves parallel to the $y$ axes.  
The existence of these snake states, which is required by topology,  
leads to the electron-hole symmetric dispersion at large $k_y$ shown in Fig. \ref{fig:bandswf}a. The opposite velocities of snake states at $\mu>0$ and $\mu<0$ 
are required by the sign reversal of $\sigma_{xy}$. 

The above topological flat bands and snake states derived from the Dirac valley at ${\bf K}_2$ have a time-reversed partner from the opposite valley at $-{\bf K}_2$ with opposite spin polarizations, 
as well as $\pi/2$-rotated copies from the two other valleys at $\pm {\bf K}_1$. 
Taking all four valleys into account, we conclude that the dislocation array causes dramatic band reconstruction on the (001) surface by 
creating flat dispersions in the vicinity of the Dirac points and a two-dimensional network of helical snake states at higher energy. 
Due to the helical nature of topological surface states, counter-propagating snake states on the same domain wall cannot backscatter, and 
the two-dimensional surface of topological crystalline insulators remains conducting even in the presence of disorder\cite{deloc}, unlike non-topological flat bands that are prone to Anderson localization.


Recently flat-band systems have attracted tremendous interest due to enhanced interaction effects associated with the high density of states and the resulting electronic instabilities. Most studies have focused on finding a completely flat band with nontrivial topology, which appears to rely on careful fine tuning of material parameters and has not been experimentally realized so far. In comparison, our work reveals a new and realistic route to achieve a partially flat band, as shown in Fig.3. For a wide range of electron or hole densities (up to about $5\times 10^{12}$cm$^{-2}$ in this example), the Fermi energy lies within the partially filled flat band. It is thus natural to ask whether interesting interaction-driven phenomena are expected to arise.  

In this regard, we note that a wide class of IV-VI semiconductor (001) multilayers and bilayers is superconducting. 
Superconductivity was first discovered in PbTe/SnTe,  PbSe/PbS, PbTe/PbSe, PbS/YbS and PbTe/YbS superlattices long ago\cite{sl1,sl2,sl3}, 
and recently found  in two-layer sandwiches of PbTe/PbS, PbTe/PbSe, and PbTe/YbS with layers 40--300nm thick\cite{bl1, bl2}. 
The transition temperatures $T_c$ are in the range of 2.5--6.4K, which is unusually high for semiconductors especially given that the individual constituent materials are non-superconducting (the only exception is SnTe with the very low $T_c$ of 0.22K). 
Further, the strong anisotropy of the upper critical field reveals that the observed superconductivity is two-dimensional\cite{bl1, bl2}. 
Based on these facts, it was concluded that the locus of superconductivity is at the interface. 

Remarkably, it was found that the appearance of superconductivity is dependent on the formation of a misfit dislocation array near the interface\cite{dis-sc, dis-sc2}.  
Samples with island-type growth, and therefore dislocations  that do not cover the whole interface only show partial superconducting transitions. For superconducting samples, the transition temperature $T_c$ was found to increase from 3K to 6K as the period of the misfit dislocation array $\lambda$ decreases from 23nm to 10nm\cite{dis-sc}. 

Previous works\cite{sl3, dis-sc} have proposed that superconductivity emerges from metallic 
interface states created by band inversion on one side of the interface where the constituent material is pseudomorphic and in compression. 
Such band inversion due to compression is only possible for narrow-gap semiconductors, 
which explains the absence of superconductivity down to $1.5$K in IV-VI multilayers 
consisting of only wide-gap semiconductors  (YbS/EuS, YbS/YbSe). 
However, this proposal does not take into account the indispensable role of dislocations in superconductivity,  
nor explain the origin of the unusually high transition temperature.  

Our work sheds new light on the interface superconductivity in IV-VI multilayers. 
The band inversion induced by compression leads to 
 the topological crystalline insulator phase\cite{hsieh}, and hence gives rise to topological surface states at the interface whose electronic properties 
 were correctly identified only recently\cite{hsieh, ando, hasan, poland, liu}.
At the same time, the dislocation array produces a periodically varying strain, which acts on these states to create topological flat bands.  In the presence of attractive interaction due to electron-phonon coupling, the high density of states associated with these flat bands dramatically increases the superconducting transition temperature. 

Our proposal of interface superconductivity from dislocation-induced flat bands provides a remarkable 
explanation of the unusual dependence of $T_c$ on the dislocation array period\cite{dis-sc}. As shown recently, the superconducting transition temperature $T_c$ in a flat-band system is linearly proportional to the area of the flat band in momentum space\cite{volovik,khodel}, which is {\it parametrically} enhanced compared to BCS theory. 
In our proposal, when the distance between dislocation lines is large, 
the  pseudo-magnetic field created by the strain field (\ref{uyy}) is largely concentrated around individual dislocation lines, thus 
the zero-energy Landau orbitals from different dislocations are spatially separated.  
In this regime, the area of the flat bands in momentum space is proportional to the reciprocal superlattice vector $2\pi/\lambda$, and thus $T_c$ increases with decreasing array period.  When $\lambda$ becomes too small, however, Landau orbitals start to overlap and band flatness gets destroyed, hence $T_c$ stops increasing.   
In effect,  the dependence of $T_c$  on the flat band degeneracy results in its non-monotonic dependence on the array period,  in agreement with the experimental observation\cite{dis-sc}. 
 
We further predict two testable signatures of flat bands in topological crystalline insulators  and their prominent role in interface superconductivity in IV-VI semiconductor multilayers.   
First, the flat bands and coexisting network of helical snake states generate a distinctive local density of states spectrum as a function of position and energy shown in Fig.3c, which can be detected in tunneling (magneto-)conductance measurements\cite{tokura}, de Haas-van Alphen or Shubnikov-de Haas oscillations. Second, the enhancement of superconductivity by flat bands ceases to work when  the flat bands become empty or filled. This leads to a drop in $T_c$ as the carrier density at the interface increases above a threshold, whose value     
depends on the strain field strength and dislocation array period and is estimated to be on the order of $10^{12}$cm$^{-2}$.   

While the superconducting transition temperatures in IV-VI heterostructures are not high on an absolute scale, 
the mechanism of flat band formation due to interface microstructures or intentional strain engineering may offer a viable route to high-temperature interface superconductivity---a subject of tremendous current interest\cite{interface, xue}. Further, our work   
opens up new directions for achieving other interesting phases in a realistic setting. 
In particular, when interactions are repulsive\cite{vishwanath, levin, roy}, novel quantum Hall states or fractional topological insulators may arise in the helical flat band we have found, at zero magnetic field. While electron repulsion is weak in IV-VI semiconductors due to their large dielectric constants, new topological crystalline insulator materials have recently been predicted/proposed in correlated electron systems such as
heavy fermion compounds\cite{dai, kai}, transition metal oxides\cite{fiete},  graphene multilayers\cite{kindermann} and anti-perovskites\cite{hsiehliufu}. These rapid and continuing developments hold promise for the physical realization of new states of matter in partially flat bands.

\section{Methods}

Here we provide a derivation of the strain-induced gauge field  (\ref{a}), or equivalently Dirac point displacement on the (001) surface Brillouin zone of topological crystalline insulators. 
The derivation is based on symmetry analysis.   
A generic in-plane strain $u_{ij}$ can be decomposed into three independent components: compression/dilation $u_{xx} + u_{yy}$, 
uniaxial stretch $u_{xx} - u_{yy}$, and shear $u_{xy} + u_{yx}$, which transform differently under crystal symmetries. 
   
Compression/dilation preserves the full symmetry of the (001) surface. In particular, the presence of two mirror planes $(110)$ and $(1\bar{1}0)$ guarantees 
the two pairs of Dirac points $\pm {\bf K}_1$ and $\pm {\bf K}_2$ lie along the mirror-invariant lines $\Gamma X_1$ and $\Gamma X_2$ respectively, in the surface Brillouin zone\cite{hsieh}. 
Importantly, the Dirac point positions on the $\Gamma X$ lines are not constrained by symmetry\cite{liu}; they vary continuously under strain\cite{serbyn}. 
As shown in recent ab-initio calculations\cite{strain1, qian}, a compressive  (tensile) strain moves $\pm {\bf K}_1$ and $\pm {\bf K}_2$ towards (away from) the Brillouin zone center $\Gamma$ (see Fig.1b). 

Uniaxial stretch  in the [1$\bar 1$0] direction  preserves both $(110)$ and $(1{\bar 1}0)$ mirror planes but breaks the 
four-fold symmetry: $u_{xx} - u_{yy}$ is odd under the $\pi/2$ rotation $x \rightarrow y, \; y \rightarrow -x$. As a consequence, the Dirac points $\pm {\bf K}_1$ and 
$\pm {\bf K}_2$ move along the $\Gamma X$ lines by an equal distance but in opposite directions: $\pm {\bf K}_1$ move inward and $\pm {\bf K}_2$ move outward (see Fig.1c).       

Shear strain  breaks both (110) and (1$\bar 1$0) mirror symmetries, as well as the four-fold rotation symmetry. 
Therefore, Dirac points move perpendicular to the $\Gamma X$ lines, and the displacement vector ${\bf K}'_1 - {\bf K}_1$ is opposite to ${\bf K}'_2 - {\bf K}_2$ after the $\pi/2$ rotation (see Fig.1d). 
  
It follows from the above analysis that a generic strain-induced gauge field ${\bf A}_j$ acting on the Dirac fermion at valley ${\bf K}_j$ 
consists of contributions from compression/dilation, stretch and shear, which involve three independent coupling constants. 
Adding up the corresponding Dirac point displacements  leads to the expression of ${\bf A}_j$ in (\ref{a}). Note that the form of ${\bf A}_j$ in topologically crystalline insulators 
is different from its counter-part in graphene, which has been extensively studied in recent years\cite{manes, guinea, neto, levy}. 
For example, unlike here, a dilation in graphene does not generate a gauge field due to the pinning of Dirac points at Brillouin zone corners.  
This difference arises from the important distinction in crystal symmetry and electronic topology. 

\section{Additional information}
Correspondence and requests for materials should be addressed to LF.

\section{Acknowledgements}
We thank Maksym Serbyn and Andrea Allais for helpful discussions, as well as Yoichi Ando and Robert Cava for valuable comments on the manuscript. 
This work is supported by DOE Office of Basic Energy Sciences,  Division of Materials Sciences and Engineering under award DE-SC0010526 (LF). 
ET acknowledges support from NSF Grants DMR-1005541, NSFC 11074140, and NSFC 11274192.

\section{Authors Contribution}
Both authors contributed to theoretical and numerical parts of the research, and the writing of the manuscript.

\end{document}